\documentclass[conference]{IEEEtran}
\IEEEoverridecommandlockouts
\usepackage{cite}
\usepackage{amsmath,amssymb,amsfonts}
\usepackage{algorithmic}
\usepackage{graphicx}
\usepackage{textcomp}
\usepackage{xcolor}
\def\BibTeX{{\rm B\kern-.05em{\sc i\kern-.025em b}\kern-.08em
    T\kern-.1667em\lower.7ex\hbox{E}\kern-.125emX}}
\begin{document}

\title{Towards speech enhancement using a\\variational U-Net architecture\\
\thanks{Funded by Deutsche Forschungsgemeinschaft (DFG, German Research Foundation) under project ID 352015383, SFB 1330/B3.}
}

\author{
\IEEEauthorblockN{Eike J. Nustede\IEEEauthorrefmark{1}, Jörn Anemüller\IEEEauthorrefmark{1}}
\IEEEauthorblockA{\IEEEauthorrefmark{1}
\textit{Carl von Ossietzky University Oldenburg, Computational Audition Group,}\\
\textit{Dept. med. Physics \& Acoustics and Cluster of Excellence Hearing4all,}\\
\textit{Oldenburg, Germany}
\\\{eike.jannik.nustede, joern.anemueller\}@uol.de}
}

\maketitle

\begin{abstract}
We investigate the viability of a variational U-Net architecture for denoising of single-channel audio data. Deep network speech enhancement systems commonly aim to estimate filter masks, or opt to work on the waveform signal, potentially neglecting relationships across higher dimensional spectro-temporal features. We study the adoption of a probabilistic bottleneck into the classic U-Net architecture for direct spectral reconstruction. Evaluation of several ablation network variants is carried out using signal-to-distortion ratio and perceptual measures, on audio data that includes known and unknown noise types as well as reverberation. Our experiments show that the residual (skip) connections in the proposed system are a prerequisite for successful spectral reconstruction, i.e., without filter mask estimation. Results show, on average, an advantage of the proposed variational U-Net architecture over its classic, non-variational version in signal enhancement performance under reverberant conditions of 0.31 and 6.98 in PESQ and STOI scores, respectively. Anecdotal evidence points to improved suppression of impulsive noise sources with the variational U-Net compared to the recurrent mask estimation network baseline.
\end{abstract}
\section{Introduction}
\label{sec:intro}
Speech in real-world environments is inevitably distorted by background noise, reverberation, or competing speakers, resulting in degraded speech intelligibility, and decreased performance in applications such as automatic speech recognition \cite{Li2014} and speaker identification \cite{Ming2007}. Speech enhancement is used to improve signal to noise ratio (SNR), thus increasing speech quality.

Recent approaches leverage non-linear modeling capabilities of deep networks and are effective in realistic environments that include non-stationary noise and varying acoustic characteristics. These methods commonly follow two approaches, generating filter masks or directly mapping the noisy mixture to enhanced speech, the latter being referred to as end-to-end systems.

Large deep networks predominantly use log-power spectra with an extended temporal context to learn features best representing clean speech \cite{Xu2015, Kumar2016}.
In contrast, the comparatively smaller auto-encoder (AE) models favor more compact features such as Mel-frequency power spectra \cite{Lu2013} and short term Fourier transform (STFT) spectra computed across short utterances \cite{Wang2014, Wang2016, Yu2020} or small temporal contexts \cite{Xia2020}.
Several state-of-the-art networks focus on using deep architectures directly on the time domain signals to generate ideal filter masks based on self-learned speech representations, e.g., \cite{Bosca2020, Luo2019}, where each mask corresponds to one target speaker.

The present work's main contribution is to propose and study the variational U-Net architecture for speech enhancement, which, to the best of our knowledge, has not been used for acoustic signal processing previously. We hypothesize that the inclusion of a probabilistic bottleneck into the U-Net architecture increases robustness towards out-of-distribution effects, such as unknown noise types or reverberation. We investigate the performance of the proposed model, as well as several ablated versions thereof.

\begin{figure}[t]
    \centerline{\includegraphics[width=6.8cm]{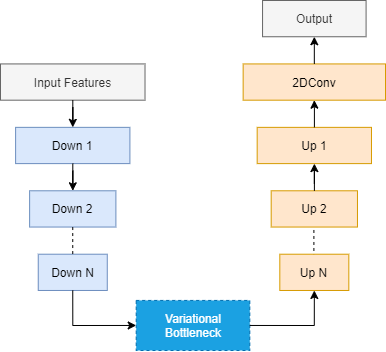}}
    \centerline{\footnotesize (a) Variational autoencoder (VAE)}\bigskip

  \centerline{\includegraphics[width=6.8cm]{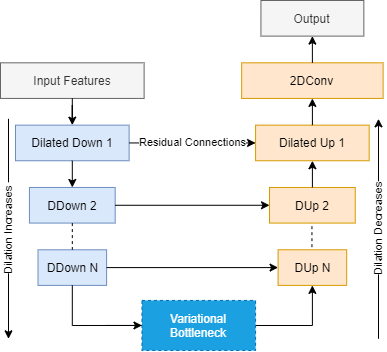}}
 \centerline{\footnotesize (b) Variational U-Net}

\caption{Model architectures of \textbf{(a)} the VAE and \textbf{(b)} the variational U-Net which includes the residual connections characteristic for U-Nets, and dilated convolutions.
\textit{N} denotes the depth of the encoder/decoder hierarchy.}
\label{fig:networks}
\end{figure}

\section{Variational U-Networks}
\label{sec:system}

\subsection{Variational U-Net Architecture}
The U-Net architecture \cite{Ronneberger2015}, originally proposed for biomedical image segmentation, has recently been used for audio source separation tasks \cite{Narayanaswamy2019, Stoller2018, Jansson2017}.
It has been adapted for speech enhancement through the addition of dilated convolutions \cite{Bosca2020}.
Filter mask generation with a dilation U-Net has been shown to perform better than without dilation, indicating the usefulness of dilation in such a network. In computer vision, the U-Net architecture has successfully been combined with the probabilistic modeling found in variational auto-encoders (VAE) for image segmentation tasks \cite{Esser2018, Kohl2018, kohl2019}.
The variational approach replaces the deterministic bottleneck with a generative Gaussian model.

We propose to combine the dilation U-Net with a VAE approach and leverage them for speech enhancement.
Additional information contained in a U-Net's residual connections, the across-frequency and -time relations found by dilated convolutions, and the robustness towards out of distribution effects of a variational approach are advantageous for real world applications. By moving in a modular fashion from the standard VAE to the proposed system (cf.\ Fig.~\ref{fig:networks}), we study the impact each step has w.r.t. perceived sound quality.

\subsection{Network Description}
\begin{figure}[h!]
    \centering
    \centerline{\includegraphics[width=0.79\linewidth]{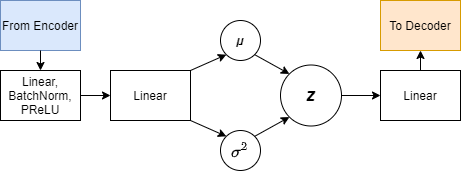}}
    \centerline{\footnotesize (a) Variational bottleneck}\bigskip
    \centerline{\includegraphics[width=0.79\linewidth]{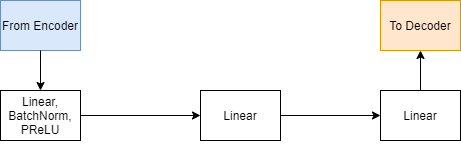}}
    \centerline{\footnotesize (b) Linear bottleneck}
    \caption{Variational and deterministic bottleneck architectures.}
    \label{fig:bottlenecks}
\end{figure}

\begin{figure}[h!]
    \centering
    \centerline{\includegraphics[width=0.79\linewidth]{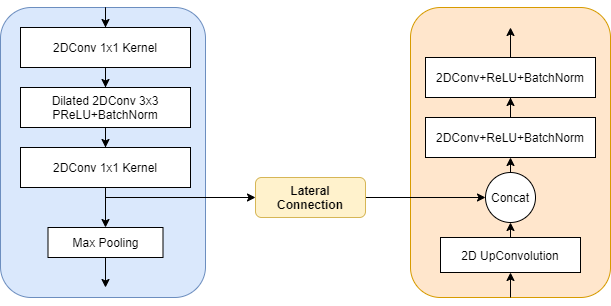}}
    \caption{Detailed view of the encoder/decoder block pairs. The encoder (left) shows the depthwise separable convolution scheme.}
    \label{fig:blocks}
\end{figure}

The general architecture of encoder and decoder follows the conventional U-Net structure, but incorporates dilated convolutions as well as a variational Bottleneck, therefore named DVUNET. The operations are consolidated into different ``Down'' and ``Up'' blocks, cf.\ Fig.~\ref{fig:networks}. Each block (Fig.~\ref{fig:blocks}) in the proposed model uses one layer of dilated convolutions implemented as depthwise separable convolutions \cite{Chollet2017} with a 3-by-3 kernel. A copy of each block's output is linked to the corresponding decoder layer, while max-pooling is applied to the original output before going into the next encoder layer. In the bottleneck (Fig.~\ref{fig:bottlenecks}), the encoded features are linearly transformed before their dimensioniality is reduced, improving training stability on our variational models. The Gaussian parameters are obtained via two linear layers. Finally, another linear layer upsamples from the bottleneck to create a tensor of the same size as the output of the encoder. The subsequent decoder blocks include transposed convolutions, that are learned upsampling operations, with 2-by-2 kernels and a stride of 2. Before each upsampling, the residuals from the encoder are concatenated to the input of a decoder block along the channel axis. As this increases channel dimension, two 3-by-3 convolutions are used to reduce the number of channels while keeping the doubled feature dimension.

Ablation models are named after their architecture, i.e., they are based on our DVUNET architecture but with selected parts removed. As such, taking away the variational bottleneck from the DVUNET is abbreviated DUNET. Further removing the dilated convolutions results in the UNET. Removing the residual connections from DVUNET creates the DVAE model, i.e., a VAE with dilations. 
Removing the variational bottleneck from the VAE results in the convolutional auto-encoder (AE).

\subsection{Loss Function Formulation}
Speech enhancement can be done either in the spectral domain or on the waveform of the acoustic signal. We chose spectral features, as it would require a deeper, larger, network to learn appropriate speech representations directly from the audio signal. Thus we define our goal as the reconstruction of a clean spectrogram from a noisy one.

Starting in the spectral domain, the input signal $X(k)$, including speech $X_s(k)$ and noise $X_n(k)$ components, is given by
\begin{equation}
    X(k) = X_s(k) + X_n(k).
\end{equation}
Let $S(k)$, $S_s(k)$ and $S_n(k)$ denote respective log-power spectra, with $S(k)$ being the input of our model and $S_s(k)$ the training target. The mean-squared error (MSE) is used as the reconstruction metric of the spectrogram and expressed as 
\begin{equation}
    L_{MSE} = E \big[ \| S_s(k) - Y(k) \|^2 \big]
\end{equation}
where $Y(k)$ denotes the enhanced spectrum at the output of the network.

The VAE related loss, the KL-Divergence, described as $D_{KL}(Q(z|S)\|P(z))$ \cite{Kingma2014}, minimizes the distance between the latent space encoding $z$ computed by the encoder $Q(z|S)$ of the model, while the decoder estimates the generative model $P(Y|z)$ under a Gaussian prior $P(z)$ assumption.

Balancing the mean-squared error against the KL-divergence with the weight $w_{KL}$  lets us formulate the overall training metric as 
\begin{equation}
    L = L_{MSE} + w_{KL} \cdot D_{KL}.
\end{equation}

\begin{figure*}[h!]
\hskip1.3cm
\begin{minipage}{.39\linewidth}
    \centering
    \centerline{\includegraphics[height=5cm]{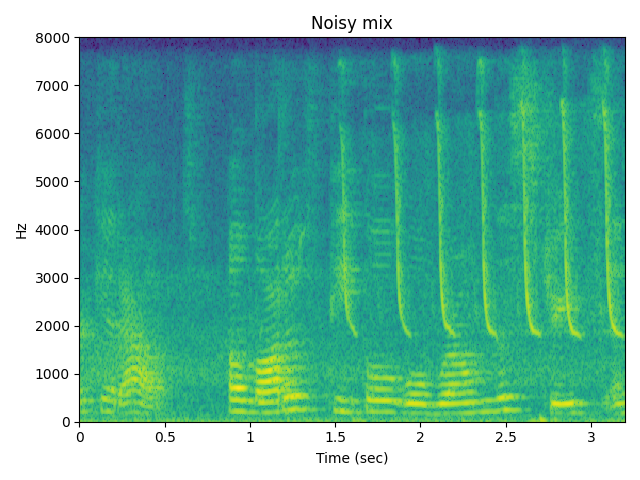}}
    \centerline{(a) Input audio spectrogram}\medskip
\end{minipage}
\hskip1cm
\begin{minipage}{.39\linewidth}
 \centering
  \centerline{\includegraphics[height=5cm]{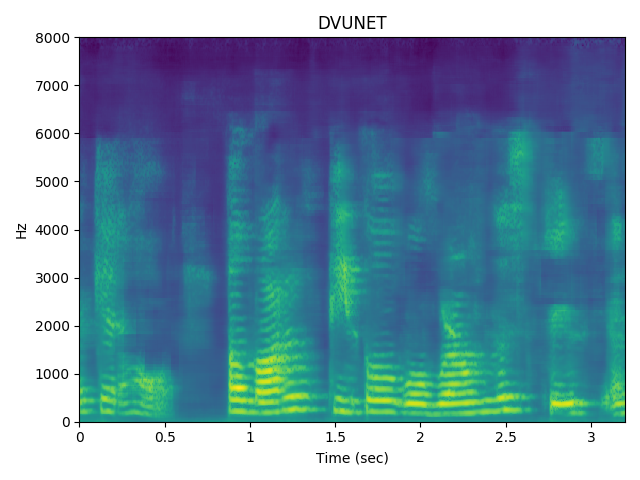}}
 \centerline{(b) DVUNET output}\medskip
\end{minipage}

\hskip1.3cm
\begin{minipage}{.39\linewidth}
    \centering
    \centerline{\includegraphics[height=5cm]{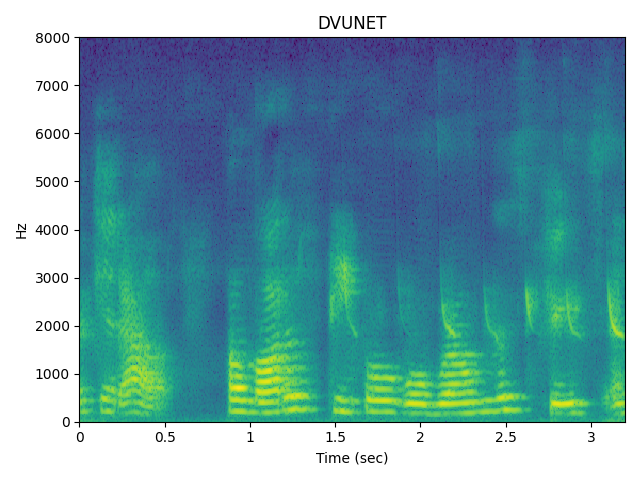}}
    \centerline{(c) Spectrogram of DVUNET enhanced signal}\medskip
\end{minipage}
\hskip1cm
\begin{minipage}{.39\linewidth}
 \centering
  \centerline{\includegraphics[height=5cm]{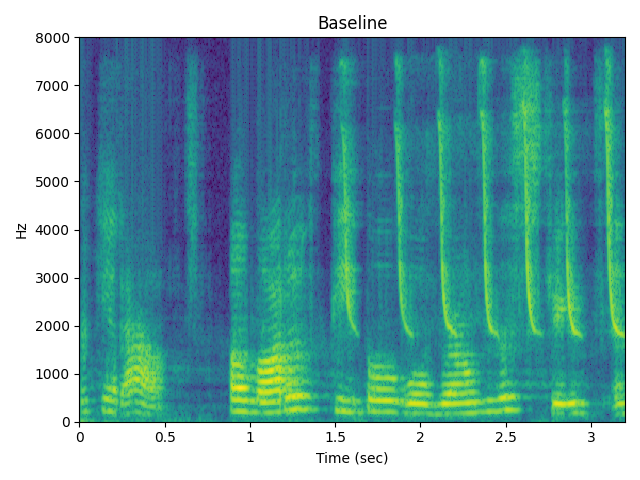}}
 \centerline{(d) Spectrogram of baseline enhanced signal}\medskip
\end{minipage}

\caption{Example of input and enhanced spectra from test data with reverberation. \textbf{(a)} Noisy, reverberant input signal. Noise dominated by bird chirping, as is  visible from about 1.5~s onwards. \textbf{(b)} Spectral reconstruction produced by variational U-Net with dilated convolutions (DVUNET), showing finer spectral structures and less noise than the final audio signal shown in (c). \textbf{(c)} Spectrogram of enhanced time-domain audio signal produced by DVUNET output together with the noisy input phase, retaining more residual chirp sounds. \textbf{(d)} Spectrogram of reconstructed audio signal enhanced by the baseline system. }
\label{fig:spectra}
\end{figure*}

\section{Experiments}
\label{sec:experiment}

\subsection{Data}
\label{sec:dataset}

\begingroup
\setlength{\tabcolsep}{5.5pt}
\begin{table}[t]
    \centering
    \caption{Performance (SI-SDR) in dependence on training set length and model characteristics (V: variational, U: U-Net, D: dilated convolutions). Identical 10h test set in all conditions.}
    \begin{tabular}{|l*{5}{|r}|}
    \hline
        \multicolumn{3}{|c}{Algorithm} & \multicolumn{3}{|c|}{SI-SDR [dB] on 10h test set}
        \\\hline 
        \multicolumn{1}{|c|}{ }&\multicolumn{1}{c|}{ }&\multicolumn{1}{c|}{ }& \multicolumn{3}{c|}{training duration}\\
        Model & Struct. & \hskip-0pt \#Param. &    1h & 15h & 100h 
        \\\hline 
        DVUNET & V+U+D  & 134 M & 14.38 $\pm$ 0.03 &   15.90 & 16.68 \\ 
        DUNET  & U+D    & 133 M & 14.39 $\pm$ 0.03 &   15.99 &   16.70 \\ 
        UNET   & U      & 146 M & 14.53 $\pm$ 0.02 &   16.08 & 16.51 \\ 
        DVAE   & V+D    &  44 M & 8.06 $\pm$ 0.11 &    8.69 & 8.84\\ 
        VAE    & V      &  56 M &  8.37 $\pm$ 0.06 &   9.16 &   9.20 \\ 
        AE     & ./.    &  56 M &  9.74 $\pm$ 0.01 &   12.77 &   13.25 \\ 
        baseline & ./.  & ./.  &  10.52           &   10.52         &   10.52 \\ 
        input data & ./.  & ./.  &  9.98          &    9.98         &    9.98 \\ 
        \hline
    \end{tabular}
    \label{tab:results}
\end{table}
\endgroup

Datasets for training and testing are generated with the scripts and audio files provided within the Deep Noise Suppression Challenge 2020 \cite{Reddy2019}. Target speech is normalized on a per utterance basis, noise added to each utterance is scaled to SNR randomly chosen from 21~SNR levels in the range 0 to 20 dB SNR. The speech and noise files are randomly selected and mixed to create noisy utterances. All audio files are 30 seconds long and sampled at 16 kHz.

Log-scaled power spectra are computed with a 1024-point STFT, Hann window of length 25~ms, and window-shift 6.25~ms. Training sets of duration 1~hour, 15~hours, and 100~hours, and a validation set of 2 hours are used during training and model selection. The test set of length 10 hours is used across all experiments, irrespective of training duration. To study the effect of reverberation during testing, we employed an additional test set of length 25~minutes with and without reverberation; data also included in the Deep Noise Suppression challenge. Note that no reverberation was present in any training set.

\subsection{Network Configuration}
\label{sec:network}
Input and output of the network are given by a spectrogram patch of length 3.2~s, representing log-scaled spectro-temporal magnitude in 512 time frames and 512 frequency bins. During training, data are presented in mini batches of 16 spectrogram patches. The model's $N=9$ encode/decoder block pairs (Fig.~\ref{fig:networks}b and Fig.~\ref{fig:blocks}) operate with channel dimensions (64, 128, 256, 512, 1024, 1024, 1024, 1024, 1024) in subsequent levels of the processing hierarchy, while simultaneously reducing feature size from ($512 \times 512$) at input/output to ($1 \times 1$) at the bottleneck in steps of factor $1/2$. Bottleneck size is $256$ with the bottleneck distribution modeled by a diagonal Gaussian with 256 mean and variance parameters. Implementation of the variational bottleneck uses the standard reparameterization approach \cite{Kingma2014}. Dilations across hierarchical levels scale with the block number, starting at 1 (no dilation) and increasing up to $N=9$.  

\subsection{Training}
\label{sec:training}
Training uses the adam optimizer with learning rate 0.001 and learning rate warm-up for the first 100 batches of the 15~h and 100~h training sets, while the 1~h training set uses a static learning rate of 0.001.
Maximum training epochs is set to 200 epochs and validation loss is evaluated ten times (100~h training set), twice (15~h training set), and once (1~h training set) per epoch, respectively.
Validation-based early stopping terminates training after 10 non-improving epochs. 
Model performance variability is determined for the 1~h training set with 10-fold cross-validation. Resulting mean and standard-error-of-the-mean values for test set performance are reported in Table \ref{tab:results}, indicating that training variability is comparably small.

\subsection{Evaluation Setup}
\label{sec:eval} 
We evaluate speech reconstruction quality in terms of the scale-invariant signal to distortion ratio (SI-SDR, \cite{Roux2019}) on the 10~h test set for all model variants and the three training set sizes as shown in table~\ref{tab:results}. The baseline model from the deep noise suppression challenge \cite{Xia2020} is based on a recurrent neural network and is available as a pre-trained model that has been trained in a mixed-SNR condition on five different SNRs in the range from 0~dB to 40~dB, i.e., no re-training of the baseline system was possible. Therefore, we note that identical baseline performance numbers are reported irrespective of the size of the training set that our models were trained on.

Objective measures of speech quality and the generalization from non-reverberant training to reverberant testing conditions are shown in table~\ref{tab:perceptive}. Speech quality was evaluated in terms of the perceptual evaluation of speech quality (PESQ) and short-time objective intelligibility (STOI) measures. Reverberant and non-reverberant test data were obtained from the 25~min.\ long reverberation test set of the deep noise suppression challenge. 

Time-domain reconstruction of audio signals for evaluation was performed by combining the log-magnitude spectrogram of the network output with the original phase of the noisy input signal spectrogram.

\section{Results \& Discussion}
\label{sec:results}

Speech reconstruction quality (SI-SDR, table~\ref{tab:results}), overall,  increases towards larger training sets, as expected for models with millions of parameters. Model architectures that include a U-Net component (UNET, DUNET, DVUNET) all perform with significantly higher SI-SDRs than those without (AE, VAE, DVAE, and baseline), implying that their residual connections are important for the task. In comparison, the differences across U-Net-type models are minor with UNET performing slightly better on the 1~h and 15~h training task, while DUNET and DVUNET perform slightly better for 100~h training. Thus, the variational bottleneck of DVUNET does not result in improved performance for this scenario, and the effect of dilated convolutions in DUNET is comparably small.

Objective perceptual measure evaluation (PESQ and STOI, table~\ref{tab:perceptive}) with evaluation in non-reverberant test conditions, similarly, shows the improved performance of U-Net-type models, with again marginal differences between UNET, DUNET and DVUNET. Ablation models without U-Net-type residual connections (DVAE, VAE, AE) do not show competitive performance in this experiment. 

In reverberant testing conditions, UNET, DUNET, and DVUNET obtain reduced perceptual scores, as expected during a train/test-scenario mismatch. However, compared to the other models, they still perform best with the variational bottleneck model (DVUNET) giving the highest perceptual scores across reverberant conditions. The reduction in performance when increasing training set length to 100~h that is observed for the dilation models might indicate overfitting, such as modeling of long-range spectro-temporal relations that are not present in the reverb test set. The variational bottleneck proves to be a beneficial architecture component, particularly under the reverberant testing scenario.

Evaluated across conditions and perceptual measures (table~\ref{tab:perceptive}, last two columns), DVUNET results in the highest performance measures, with DUNET as the second-best model. The other ablation models, as well as the baseline model, produce on average clearly lower perceptual scores. We conclude that dilated convolutions with their increased spectral and temporal context, and the variational bottleneck are important model components when generalization of trained models to reverberant test conditions is required.

\begingroup
\setlength{\tabcolsep}{5.5pt}
\begin{table*}[t]
    \centering
    \caption{Perceptual evaluation on the 25--minute long test sets for 15h and 100h training data.}
    \begin{tabular}{|l|*{10}{r|}}
        \hline
        \multicolumn{1}{|c}{ }&\multicolumn{2}{c|}{no-reverb test} & \multicolumn{2}{c|}{reverb test} & \multicolumn{2}{c|}{no-reverb test} & \multicolumn{2}{c|}{reverb test} & \multicolumn{2}{c|}{}\\
        \multicolumn{1}{|c}{ }&\multicolumn{2}{c|}{15h training} & \multicolumn{2}{c|}{15h training} &\multicolumn{2}{c|}{100h training} & \multicolumn{2}{c|}{100h training} & 
        \multicolumn{2}{c|}{average}\\
        \hline
        Model & PESQ & STOI  & PESQ & STOI  & PESQ & STOI  & PESQ & STOI & PESQ & STOI \\
        \hline
        DVUNET  & 2.99 & 93.74 & \textbf{2.44} & 79.30 & \textbf{3.14} & \textbf{94.85}  & \textbf{1.99} & \textbf{70.93} & \textbf{2.64} & \textbf{84.71} \\ 
        DUNET  & 2.96 & \textbf{94.21} & 2.28 & \textbf{79.62} &\textbf{3.14} & 94.60 & 1.68 & 62.47 & 2.52 & 82.73\\ 
        UNET  & \textbf{3.01} & 94.08 & 1.60 & 61.01 & 3.11 & 94.60 & 1.61 & 61.22 & 2.33 & 77.73 \\ 
        DVAE  & 1.25 & 48.33  & 1.29 & 33.43 & 1.27 & 48.67 & 1.30 & 35.20 & 1.28 & 41.41 \\ 
        VAE     & 1.25  & 49.53  & 1.40 & 35.88 & 1.32 & 50.95 & 1.31 & 37.36 & 1.32 & 43.43 \\ 
        AE     & 1.40  & 71.73  & 1.46 & 60.13  & 1.55 & 77.01  & 1.47 & 56.54 & 1.47 & 66.35 \\ 
        baseline & 2.42 & 77.62 & 2.14 & 69.24  & 2.42 & 77.62 & 2.14 & 69.24 & 2.28 & 73.43 \\
        input data  & 2.52 & 91.51  & 2.16 & 86.62  & 2.52 & 91.51 & 2.16 & 86.62 & 2.34 & 89.07 \\
        \hline
    \end{tabular}
    \label{tab:perceptive}
\end{table*}
\endgroup

Fig.~\ref{fig:spectra} illustrates an example where the DVUNET algorithm suppresses wide-band background noise as well as chirp sounds with confined spectro-temporal support.
We note that the slight ``blurring'' in the DVUNET output is inaudible and that in informal subjective listening, the DVUNET reconstructed speech signal is perceived as high quality without artefacts and a low degree of remaining noise compared to baseline.
Panel (c) highlights that some audible noise enters the reconstructed signal through effects of the noisy phase, indicating that phase-aware processing may further improve results.

\section{Conclusion}
The present contribution developed a U-Net architecture with a variational bottleneck and performed speech enhancement with it. To our knowledge, a variational U-Net has not been proposed for acoustic signals before, with related variants having been developed for image segmentation \cite{Esser2018, Kohl2018, kohl2019}.
The variational U-Net and ablation systems thereof were compared using data and baseline system from the deep noise separation challenge. Results indicate that the variational bottleneck is important for generalization from non-reverberant training to reverberant test conditions. Necessity of lateral U-Net connections and usefulness of dilated convolutions have also been corroborated by results.

The observation that the variational model performs best is consistent with the hypothesis that the generative Gaussian model in the variational bottleneck is better at steering the reconstruction towards output data that is close to the training distribution. Purely deterministic models, including the standard (non-variational) U-Net architecture, have been shown in our experiments to perform worse under train/test-mismatch.

Future work will aim to further investigate applications with a distribution shift between model and observation. Dereverberation, e.g., may also benefit from a variational approach to source reconstruction. 

\bibliographystyle{./IEEEtran}
\bibliography{paper_bib_v2}

\begin{thebibliography}{10}
\providecommand{\url}[1]{#1}
\csname url@samestyle\endcsname
\providecommand{\newblock}{\relax}
\providecommand{\bibinfo}[2]{#2}
\providecommand{\BIBentrySTDinterwordspacing}{\spaceskip=0pt\relax}
\providecommand{\BIBentryALTinterwordstretchfactor}{4}
\providecommand{\BIBentryALTinterwordspacing}{\spaceskip=\fontdimen2\font plus
\BIBentryALTinterwordstretchfactor\fontdimen3\font minus
  \fontdimen4\font\relax}
\providecommand{\BIBforeignlanguage}[2]{{%
\expandafter\ifx\csname l@#1\endcsname\relax
\typeout{** WARNING: IEEEtran.bst: No hyphenation pattern has been}%
\typeout{** loaded for the language `#1'. Using the pattern for}%
\typeout{** the default language instead.}%
\else
\language=\csname l@#1\endcsname
\fi
#2}}
\providecommand{\BIBdecl}{\relax}
\BIBdecl

\bibitem{Li2014}
J.~Li, L.~Deng, Y.~Gong, and R.~Haeb-Umbach, ``An overview of noise-robust
  automatic speech recognition,'' \emph{IEEE/ACM Transactions on Audio, Speech,
  and Language Processing}, vol.~22, pp. 745--777, 2014.

\bibitem{Ming2007}
J.~Ming, T.~J. Hazen, J.~R. Glass, and D.~A. Reynolds, ``{Robust speaker
  recognition in noisy conditions},'' \emph{IEEE Trans. Audio, Speech Lang.
  Process.}, vol.~15, no.~5, pp. 1711--1723, 2007.

\bibitem{Xu2015}
Y.~Xu, J.~Du, L.~R. Dai, and C.~H. Lee, ``{A regression approach to speech
  enhancement based on deep neural networks},'' \emph{IEEE/ACM Trans. Audio
  Speech Lang. Process.}, vol.~23, no.~1, pp. 7--19, 2015.

\bibitem{Kumar2016}
A.~Kumar and D.~Florencio, ``{Speech enhancement in multiple-noise conditions
  using deep neural networks},'' in \emph{INTERSPEECH}, 2016, pp. 3738--3742.

\bibitem{Lu2013}
X.~Lu, Y.~Tsao, S.~Matsuda, and C.~Hori, ``{Speech enhancement based on deep
  denoising autoencoder},'' in \emph{INTERSPEECH}, 2013, pp. 436--440.

\bibitem{Wang2014}
Y.~Wang, A.~Narayanan, and D.~L. Wang, ``{On training targets for supervised
  speech separation},'' \emph{IEEE/ACM Trans. Audio Speech Lang. Process.},
  vol.~22, no.~12, pp. 1849--1858, 2014.

\bibitem{Wang2016}
S.~S. Wang, H.~T. Hwang, Y.~H. Lai, Y.~Tsao, X.~Lu, H.~M. Wang, and B.~Su,
  ``{Improving denoising auto-encoder based speech enhancement with the speech
  parameter generation algorithm},'' in \emph{2015 Asia-Pacific Signal Inf.
  Process. Assoc. Annu. Summit Conf. APSIPA ASC}, 2015, pp. 365--369.

\bibitem{Yu2020}
\BIBentryALTinterwordspacing
C.~Yu, R.~E. Zezario, J.~Sherman, Y.-Y. Hsieh, X.~Lu, H.-M. Wang, and Y.~Tsao,
  ``{Speech Enhancement based on Denoising Autoencoder with Multi-branched
  Encoders},'' pp. 1--12, 2020. [Online]. Available:
  \url{http://arxiv.org/abs/2001.01538}
\BIBentrySTDinterwordspacing

\bibitem{Xia2020}
Y.~Xia, S.~Braun, C.~K. Reddy, H.~Dubey, R.~Cutler, and I.~Tashev, ``{Weighted
  Speech Distortion Losses for Neural-Network-Based Real-Time Speech
  Enhancement},'' in \emph{ICASSP, IEEE Int. Conf. Acoust. Speech Signal
  Process.}, 2020, pp. 871--875.

\bibitem{Bosca2020}
\BIBentryALTinterwordspacing
A.~Bosca, A.~Gu{\'{e}}rin, L.~Perotin, and S.~Kiti{\'{c}}, ``{Dilated U-net
  based approach for multichannel speech enhancement from First-Order
  Ambisonics recordings},'' pp. 1--5, 2020. [Online]. Available:
  \url{http://arxiv.org/abs/2006.01708}
\BIBentrySTDinterwordspacing

\bibitem{Luo2019}
Y.~Luo and N.~Mesgarani, ``{Conv-TasNet: Surpassing Ideal Time-Frequency
  Magnitude Masking for Speech Separation},'' \emph{IEEE/ACM Trans. Audio
  Speech Lang. Process.}, vol.~27, no.~8, pp. 1256--1266, 2019.

\bibitem{Ronneberger2015}
O.~Ronneberger, P.~Fischer, and T.~Brox, ``{U-Net: Convolutional networks for
  biomedical image segmentation},'' \emph{Lect. Notes Comput. Sci.}, vol. 9351,
  pp. 234--241, 2015.

\bibitem{Narayanaswamy2019}
\BIBentryALTinterwordspacing
V.~S. Narayanaswamy, S.~Katoch, J.~J. Thiagarajan, H.~Song, and A.~Spanias,
  ``{Audio Source Separation via Multi-Scale Learning with Dilated Dense
  U-Nets},'' 2019, arXiv:1904.04161. [Online]. Available:
  \url{http://arxiv.org/abs/1904.04161}
\BIBentrySTDinterwordspacing

\bibitem{Stoller2018}
D.~Stoller, S.~Ewert, and S.~Dixon, ``{Wave-U-Net: A multi-scale neural network
  for end-to-end audio source separation},'' in \emph{ISMIR, Proc. 19th Int.
  Soc. Music Inf. Retr. Conf.}, 2018, pp. 334--340.

\bibitem{Jansson2017}
A.~Jansson, E.~Humphrey, N.~Montecchio, R.~Bittner, A.~Kumar, and T.~Weyde,
  ``{Singing Voice Separation with Deep U-Net CNN},'' in \emph{ISMIR, Proc.
  19th Int. Soc. Music Inf. Retr. Conf.}, 2017.

\bibitem{Esser2018}
P.~Esser, E.~Sutter, and B.~Ommer, ``{A Variational U-Net for Conditional
  Appearance and Shape Generation},'' in \emph{CVPR, IEEE Comput. Soc. Conf.
  Comput. Vis. Pattern Recognit.}, 2018, pp. 8857--8866.

\bibitem{Kohl2018}
S.~A.~A. Kohl, B.~Romera-Paredes, C.~Meyer, J.~{De Fauw}, J.~R. Ledsam, K.~H.
  Maier-Hein, S.~M. {Ali Eslami}, D.~J. Rezende, and O.~Ronneberger, ``{A
  probabilistic {U-Net} for segmentation of ambiguous images},'' in
  \emph{NeurIPS, Adv. Neural Inf. Process. Syst.}, 2018, pp. 6965--6975.

\bibitem{kohl2019}
\BIBentryALTinterwordspacing
S.~A.~A. Kohl, B.~Romera-Paredes, K.~H. Maier-Hein, D.~J. Rezende, S.~M.~A.
  Eslami, P.~Kohli, A.~Zisserman, and O.~Ronneberger, ``A hierarchical
  probabilistic u-net for modeling multi-scale ambiguities,'' 2019. [Online].
  Available: \url{http://arxiv.org/abs/1905.13077}
\BIBentrySTDinterwordspacing

\bibitem{Chollet2017}
F.~Chollet, ``{Xception: Deep learning with depthwise separable
  convolutions},'' in \emph{CVPR, IEEE Conf. Comput. Vis. Pattern Recognition},
  2017, pp. 1800--1807.

\bibitem{Kingma2014}
D.~P. Kingma and M.~Welling, ``{Auto-encoding variational bayes},'' \emph{2nd
  Int. Conf. Learn. Represent. ICLR 2014 - Conf. Track Proc.}, no.~Ml, pp.
  1--14, 2014.

\bibitem{Reddy2019}
C.~K.~A. Reddy, E.~Beyrami, J.~Pool, R.~Cutler, S.~Srinivasan, and J.~Gehrke,
  ``{A scalable noisy speech dataset and online subjective test framework},''
  in \emph{INTERSPEECH}, 2019, pp. 1816--1820.

\bibitem{Roux2019}
J.~L. Roux, S.~Wisdom, H.~Erdogan, and J.~R. Hershey, ``{SDR - Half-baked or
  Well Done?}'' in \emph{ICASSP, IEEE Int. Conf. Acoust. Speech Signal
  Process.}, 2019, pp. 626--630.

\end{thebibliography}
\end{document}